\documentclass[a4paper,12pt,dvips,oneside]{article}
\usepackage{amsmath,booktabs,dcolumn,graphicx}
\usepackage{mathptm,overcite,setspace,xspace}
\usepackage[scanall]{psfrag}

\renewcommand\baselinestretch{1.3}
\newcolumntype{d}{D{.}{.}{-1}}
\newcolumntype{L}{>{$}l<{$}}
\newcolumntype{R}{>{$}r<{$}}
\newcolumntype{C}{>{$}c<{$}}

\newcommand{\eq}[1]{Eq.~(\ref{eq:#1})}

\newcommand{\fig}[1]{Figure \ref{fig:#1}}
\newcommand{\figs}[1]{Figures \ref{fig:#1}}
\newcommand{\paper}[1]{Ref.~\citen{#1}}
\newcommand{\tab}[1]{Table \ref{tab:#1}}

\newcommand{\sect}[1]{Section \ref{sect:#1}}

\newcommand{\half}{{\textstyle\frac{1}{2}}}
\newcommand{\ie}{i.e., }

\newcommand{\K}{\,{\rm K}}

\newcommand{\Go}{G\=o\xspace}
\newcommand{\fsz}{\footnotesize}
\newcommand{\ssz}{\scriptsize}

\begin{document}

\title{Energy Landscape of a Model Protein}
\author{Mark A.~Miller and David J.~Wales \\
  {\it\small University Chemical Laboratories,
   Lensfield Road, Cambridge CB2 1EW, UK} }

\maketitle
\bigskip

\renewcommand{\baselinestretch}{1.8}
\normalsize
\section*{Abstract}
The potential energy surface of an off-lattice model protein is characterized
in detail by constructing a disconnectivity graph and by examining the
organisation of pathways on the surface. The results clearly
reveal the frustration exhibited by this system and explain why it
does not fold efficiently to the global potential energy minimum.
In contrast, when the frustration is removed by constructing a `\Go-type'
model, the resulting graph exhibits the characteristics expected for
a folding funnel.

\section{Introduction}

The potential energy surface (PES) of an interacting system determines its
structural, dynamic, and thermodynamic properties. Formally, the links between
the PES and these properties are fully defined by the stationary points on the
PES, its gradient (which gives the forces on the particles), and the partition
function. However, it is only relatively recently that explicit connections
have been sought between the overall structure of the PES,
or potential energy `landscape', and the behaviour of the system it describes.
This approach promises to provide insight into a number of fields,
including protein folding, global optimization and glass formation.
\par
In the present contribution we provide a global characterization of the
PES for a model heteropolymer, and show how this picture explains the
dynamical properties observed in previous simulations. In the original
model `frustration' prevents efficient relaxation to the global
potential energy minimum. However, when the frustration is removed by
constructing the corresponding `\Go-like' model, the deep traps disappear and
the resulting surface resembles a funnel. The term frustration was first used
in the context of spin glasses,\cite{Toulouse77a} where it is impossible to
satisfy all favourable interactions simultaneously. Analogous effects exist
in proteins:\cite{Bryngelson95a} a three-dimensional structure that brings
together two mutually attractive residues may involve generating unfavourable
contacts elsewhere (`energetic frustration'), and the interconversion of
two similar structures may require the disruption of existing favourable
interactions (`geometric frustration').
\par
The major difficulty in providing a fundamental explanation of structure,
dynamics and thermodynamics in terms of the underlying potential energy 
surface is that the number of stationary points grows very rapidly with 
the size of the system.\cite{Stillinger99a} This growth is, in fact, the
basis of Levinthal's `paradox',\cite{Levinthal69a} which points out the
apparent impossibility of a protein finding its biologically active state
in a random search amongst the astronomical number of available structures.
Some attempts to resolve the paradox proposed a reduction in the search
space from the full configuration space\cite{Levinthal68a,Camacho93a,Sali94a,Socci94a}.
Although it seems unlikely that this reduction is the solution to the
paradox, there is an implicit realization in such approaches that, in some
way, the search is not random. In terms of the energy landscape there are
two reasons for this. Firstly, conformations have different statistical
weights in the thermodynamic ensemble, and secondly, they are not arranged
at random in configuration space. Levinthal's analysis assumes that the
energy landscape is flat, like a golf course with a single hole corresponding
to the native state.\cite{Bryngelson95a} By constructing a simple model that
includes an energetic bias towards the native structure, it can be shown that
the search time on the full conformational space is dramatically reduced to
physically meaningful scales.\cite{Zwanzig92a,Zwanzig95a} 
\par
One of the first studies
to consider more explicitly the organization of the energy landscape was
that of Leopold, Montal, and Onuchic.\cite{Leopold92a} These authors proposed
that the landscape of a natural protein consists of a collection of convergent
kinetic pathways that lead to a unique native state which is thermodynamically
the most stable. Such a landscape structure was termed a `folding funnel' because
it focuses the manifold misfolded states towards the correct target. This approach
highlights the fundamental fallacy of the random search in Levinthal's `paradox'.
\par
Funnel theory has gained widespread acceptance through its development by
Wolynes and coworkers in terms of a {\em free} energy
landscape.\cite{Bryngelson95a,Wolynes95a} 
The funnel can be described in terms of the free energy gradient towards
the native structure, and the roughness---a measure of the barrier heights between
local free energy minima, which can act as kinetic traps. Folding is encouraged when
the roughness is not large compared with the energy gradient. Simulations have shown
that the folding ability can be measured by the ratio of the folding temperature,
$T_{\rm f}$, where the native state becomes thermodynamically the most stable,
to the glass transition temperature, $T_{\rm g}$, where the kinetics slow down
dramatically because of the free energy barriers.\cite{Socci94a,Socci96a} $T_{\rm g}$
is usually defined as the temperature at which the folding time passes through a certain
threshold. Folding is easiest for large $T_{\rm f}/T_{\rm g}$, since the native
state is then statistically populated at temperatures where it is kinetically accessible.
The effect of frustration is to increase the roughness of the energy landscape
relative to its gradient towards the native structure, thereby hindering relaxation
to the latter.
\par
We have recently shown\cite{Wales98c,Doye99b} how a new visualization
of the potential energy surface using disconnectivity graphs\cite{Becker97a}
reveals the features which determine relaxation of clusters
to their global potential energy minimum.
This approach has already been used by others
to examine the energy landscape of a tetrapeptide\cite{Becker97a,Czerminski90a}
and to study the effects of conformational constraints in hexapeptides\cite{Levy98a}
employing an all-atom model. In the present contribution we analyse a
coarse-grained representation of a larger polypeptide
with 46 residues. Connected sequences of minima have been
reported before for this system\cite{Berry97a}, and we will show how the disconnectivity
graph approach provides a clearer picture of the relation between the energy landscape
and dynamics.

\section{The Model Potential}

Intermediate in detail between lattice and all-atom models of proteins are
continuum bead models, in which each monomer is represented by a single bead on a chain.
These off-lattice systems have received relatively little attention in terms
of landscape analysis, but provide a useful medium for such an approach, since
atomistic representations of proteins are computationally demanding.
\par
Here we examine the effects of frustration in a model heteropolymer
introduced by Honeycutt and Thirumalai.\cite{Honeycutt90a,Honeycutt92a} These
authors proposed a `metastability hypothesis' that a polypeptide
may adopt a variety of metastable folded conformations with similar structural
characteristics but different energies. The particular state reached in the
folding process depends on the initial conditions. We shall see that this scenario
arises from frustration effects intrinsic to the model, which are not expected
for a `good folder'.
\par
The heteropolymer has $N=46$ beads linked by stiff bonds. There are three types
of bead: hydrophobic (B), hydrophilic (L), and neutral (N), and the sequence is
\begin{displaymath}
{\rm B}_9{\rm N}_3({\rm LB})_4{\rm N}_3{\rm B}_9{\rm N}_3({\rm LB})_5{\rm L}.
\end{displaymath}
The potential energy is given by\cite{Honeycutt92a}
\begin{multline}
V=\half K_r\sum_{i=1}^{N-1}(r_{i,i+1}-r_{\rm e})^2
 +\half K_\theta\sum_i^{N-2}(\theta_i-\theta_{\rm e})^2 \\
 +\epsilon\sum_i^{N-3}\left[A_i(1+\cos\varphi_i)+B_i(1+\cos3\varphi_i)\right] \\
 +4\epsilon\sum_{i=1}^{N-2}\sum_{j=i+2}^N C_{ij}\left[\left(\frac{\sigma}{r_{ij}}\right)^{12}
    -D_{ij}\left(\frac{\sigma}{r_{ij}}\right)^6\right],
\label{eq:barrelpot}
\end{multline}
where $r_{ij}$ is the separation of beads $i$ and $j$.
The first term represents the bonds linking successive beads. The bond lengths
were constrained at $r_{\rm e}$ in \paper{Honeycutt92a}, but here we follow Berry
et al.\cite{Berry97a} by replacing these constraints with stiff springs:
$K_r=231.2\,\epsilon\sigma^{-2}$, where $\sigma$ and $\epsilon$ are the units of length
and energy defined by the last term in \eq{barrelpot}. To put the energy parameter
in a physical context, the value of $\epsilon$ suggested by Berry et al.\cite{Berry97a}
is $121\K$, such as might be used for the van der Waals interactions between argon atoms.
The second term in \eq{barrelpot}
is a sum over the bond angles, $\theta_i$, defined by the triplets
of atomic positions ${\bf r}_i$ to ${\bf r}_{i+2}$, with
$K_\theta=20\,\epsilon\,{\rm rad}^{-2}$ and $\theta_{\rm e}=105^\circ$. The third term
is a sum over the dihedral angles, $\varphi_i$, defined by the quartets ${\bf r}_i$ to
${\bf r}_{i+3}$. If the quartet involves no more than one N monomer then $A_i=B_i=1.2$, generating
a preference for the trans conformation ($\varphi_i=180^\circ$), whereas if two or three
N monomers are involved then $A_i=0$ and $B_i=0.2$. This choice makes the three neutral
segments of the chain flexible and likely to accommodate turns. The last term in \eq{barrelpot}
represents the non-bonded interactions, and $\sigma$ is set equal to $r_{\rm e}$.
The coefficients for the various combinations of monomer types are as follows.
\begin{alignat*}{3}
i,j\in{\rm B}& \qquad &C_{ij}=1 \quad &D_{ij}=1 \\
i\in{\rm L},\ j\in{\rm L,B}& \qquad &C_{ij}={\textstyle\frac{2}{3}} \quad &D_{ij}=-1 \\
i\in{\rm N},\ j\in{\rm L,B,N}& \qquad &C_{ij}=1 \quad &D_{ij}=0,
\end{alignat*}
with $C_{ij}=C_{ji}$ and $D_{ij}=D_{ji}$. Hence, hydrophobic monomers experience a mutual
van der Waals attraction, and all other combinations are purely repulsive, with interactions
involving a hydrophilic monomer being of longer range.
\par
\begin{figure}[ht]
\begin{center}
\includegraphics[width=85mm]{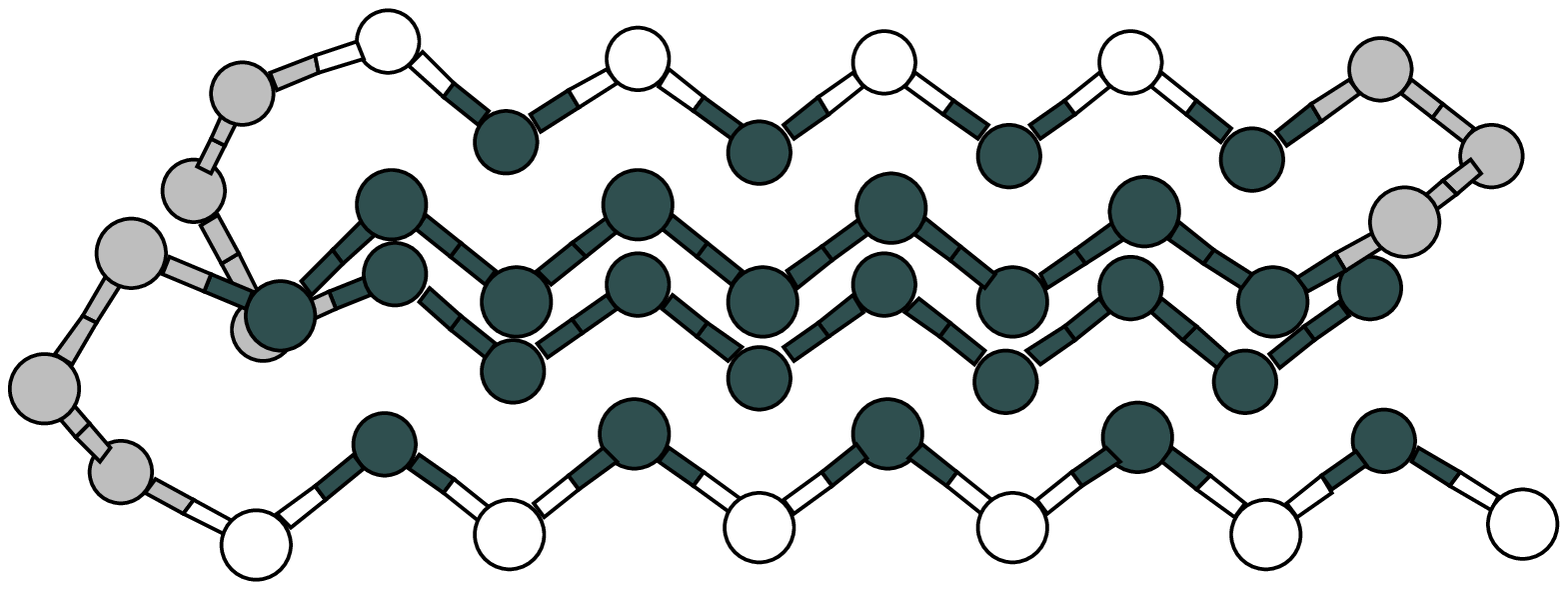}
\end{center}
\begin{center}
\includegraphics[width=43mm]{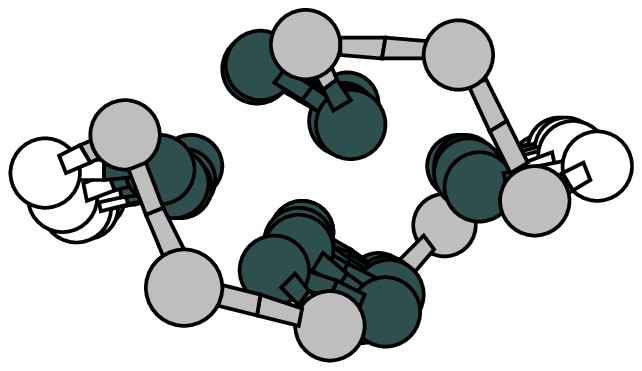}
\end{center}
\caption{
Side and end views of the global minimum of the BLN model.
Hydrophobic, hydrophilic, and neutral beads are shaded dark grey, white and
light grey, respectively.
}
\label{fig:p46}
\end {figure}

The global minimum of this system, which we call the BLN model, is a four-stranded
$\beta$-barrel,\cite{Honeycutt92a} illustrated in \fig{p46}. The hydrophobic segments
congregate at the core, and there are turns at the neutral segments. By cutting the
sequence at these regions, Vekhter and Berry have also used this model to study the
self-assembly of the $\beta$-barrel from the separated strands.\cite{Vekhter99a}

\section{Characterizing the Energy Landscape}

The most important points on a PES are the minima and the transition states that
connect them. A transition state is a stationary point at which the Hessian matrix
has exactly one negative eigenvalue whose eigenvector corresponds to the reaction
coordinate. Minima linked by higher-index saddles (the index being the number of
negative Hessian eigenvalues) must also be linked by one or more true transition
states of lower energy.\cite{Murrell68a} The pattern of stationary points and their
connectivities define the topology of the PES.

\subsection{Exploring the Landscape\label{sect:explore}}

All the transition states in the present work were located by 
eigenvector-following\cite{Pancir74a,Cerjan81a,Simons83a,Banerjee85a,Baker86a,Wales94a},
where the energy is maximized along one direction and simultaneously minimized in all
the others. Details of our implementation have been given before\cite{Wales96c}.
The minima connected to a given transition state are defined by the end points
of the two steepest-descent paths commencing parallel and antiparallel to the
transition vector (\ie the Hessian eigenvector whose eigenvalue is negative)
at the transition state. Rather than steepest-descent minimization, we have
employed a conjugate-gradient method (using only first derivatives of the potential)
to calculate the pathways. This technique gives
similar results, and has the advantage of being much faster. However, it is possible
for conjugate-gradient minimization to converge to a stationary point of higher index
than a minimum. To guard against this problem, each optimization was followed
by reoptimization with eigenvector-following to a local minimum. In the majority
of cases, the reoptimisation converged in a few steps, indicating that the
conjugate-gradient method had indeed found a true minimum.
\par
A number of similar approaches have been developed for
systematically exploring a PES by hopping between potential
wells,\cite{Tsai93b,Doye97a,Barkema96a,Mousseau98a} and these can be
adapted to obtain a topographical database in several ways.
Here we want to explore the energy landscape thoroughly, working from the
global minimum upwards. In our scheme, we commenced at the lowest-energy
known minimum, and performed an eigenvector-following search for a transition
state along the eigenvector with the smallest non-zero eigenvalue. Having
located a transition state, the connected minima were found by evaluating the
path as described above. The process was then repeated,
always starting at the lowest-energy minimum found so far, and searching
along eigenvectors in both directions in order of increasing eigenvalue.
To enable the search to explore away from the starting minimum, an
upper limit, $n_{\rm ev}$, was imposed on the number of eigenvectors to
be searched from each minimum. When $n_{\rm ev}$ eigenvectors had been exhausted,
the search moved onto the next-lowest energy minimum. We note that, even if
$n_{\rm ev}$ is set to its maximum value of $3N-6$, there is no guarantee of
finding all the transition states connected to a given minimum.
\par
The low-energy regions of the BLN model energy landscape were explored using
$n_{\rm ev}=10$ until 250 minima had been found. Because of the harmonic
bond potential, following normal modes uphill in energy did not always lead to
a transition state in a reasonable number of iterations in this system. To
compensate for this problem, the value of $n_{\rm ev}$ was raised to 20 and the
search continued until a total of 500 minima had been found. The final number of
transition states was 636.

\subsection{Visualization\label{sect:visualisation}}

A useful visual representation of a PES is provided by the disconnectivity
graph of Becker and Karplus.\cite{Becker97a} This technique was first introduced to
interpret a structural database of the tetrapeptide isobutyryl-(ala)$_3$-NH-methyl,
produced by Czerminski and Elber,\cite{Czerminski90a} and was subsequently applied
to study the effects of conformational constraints in hexapeptides.\cite{Levy98a}
The method is formally expressed\cite{Becker97a} in the language of graph theory,
but can easily be summarized as follows.
\par
At a given total energy, $E$, the minima can be grouped into disjoint sets,
called `super-basins', whose members are mutually accessible at that energy.
In other words, each pair of minima in a super-basin are connected directly or
through other minima by a path whose energy never exceeds $E$,
but would require more energy to reach a minimum in another super-basin.
At low energy there is just one super-basin---that containing the global minimum.
At successively higher energies, more super-basins come into play as new
minima are reached. At still higher energies, the super-basins coalesce as higher
barriers are overcome, until finally there is just one containing
all the minima (provided there are no infinite barriers). 
\par
A disconnectivity graph is constructed by performing the super-basin
analysis at a series of energies, plotted on a vertical scale. At
each energy, a super-basin is represented by a node, with lines joining
nodes in one level to their daughter nodes in the level below. The
choice of the energy levels is important; too wide a spacing and
no topological information is left, whilst too close a spacing
produces a vertex for every transition state and hides the overall
structure of the landscape. The horizontal position of the
nodes is arbitrary, and can be chosen for clarity.
In the resulting graph, all branches terminate at local minima, while
all minima connected directly or indirectly to a node are mutually
accessible at the energy of that node.
\par
Visualization of the PES in terms of connectivity patterns between minima represents
a mapping from the full configuration space onto the underlying `inherent
structures'\cite{Stillinger82a}. Although this approach discards information about
the volume of phase space associated with each minimum, the density of minima with
energy can provide a qualitative impression of the volumes associated with the various
regions of the energy landscape.
\par
\begin{figure}[ht]
\psfrag{(a)}{\fsz (a)}
\psfrag{(b)}{\fsz (b)}
\psfrag{(c)}{\fsz (c)}
\centerline{\includegraphics[width=70mm]{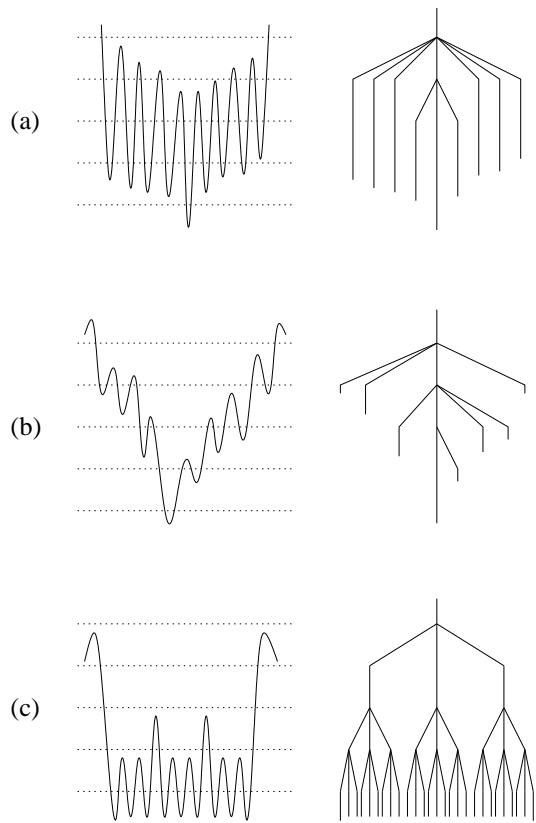}}
\caption{
Schematic examples of potential energy surfaces (potential energy as a
function of some generalized coordinate) and the corresponding
disconnectivity graphs. In each case, the dotted lines indicate the
energy levels at which the super-basin analysis has been made.
(a) A gently sloping funnel with high barriers, (b) a steeper funnel
with lower barriers, and (c) a `rough' landscape.
}
\label{fig:demo}
\end{figure}

Some example schematic potential energy surfaces and the corresponding disconnectivity
graphs are illustrated in \fig{demo}. The first two examples demonstrate that
a funnel-shaped landscape produces a disconnectivity graph with a single stem
leading to the global minimum, from which branches sprout corresponding to
local minima that are progressively cut off as the energy descends below the
barriers. The contrasting nature of the funnels in \figs{demo}(a) and (b)
is immediately discernible from the corresponding graphs, where we see that
the higher barriers and lower potential energy gradient towards the global minimum in
(a) produce long dangling branches in the disconnectivity graph. \fig{demo}(c)
is qualitatively different. The PES possesses a hierarchical arrangement of
barriers, giving rise to multiple sub-branching in the graph. The strength of
the disconnectivity graph in representing the topology of the PES is that it
is independent of the dimensionality of the system, whereas schematic
plots of the potential energy itself are restricted to one or two dimensions.
\par
\begin{figure}[ht]
\psfrag{-36}[r][r]{\fsz $-36$}
\psfrag{-38}[r][r]{\fsz $-38$}
\psfrag{-40}[r][r]{\fsz $-40$}
\psfrag{-42}[r][r]{\fsz $-42$}
\psfrag{-44}[r][r]{\fsz $-44$}
\psfrag{-46}[r][r]{\fsz $-46$}
\psfrag{-48}[r][r]{\fsz $-48$}
\psfrag{-50}[r][r]{\fsz $-50$}
\psfrag{-52}[r][r]{\fsz $-52$}
\psfrag{-54}[r][r]{\fsz $-54$}
\centerline{\includegraphics[width=85mm]{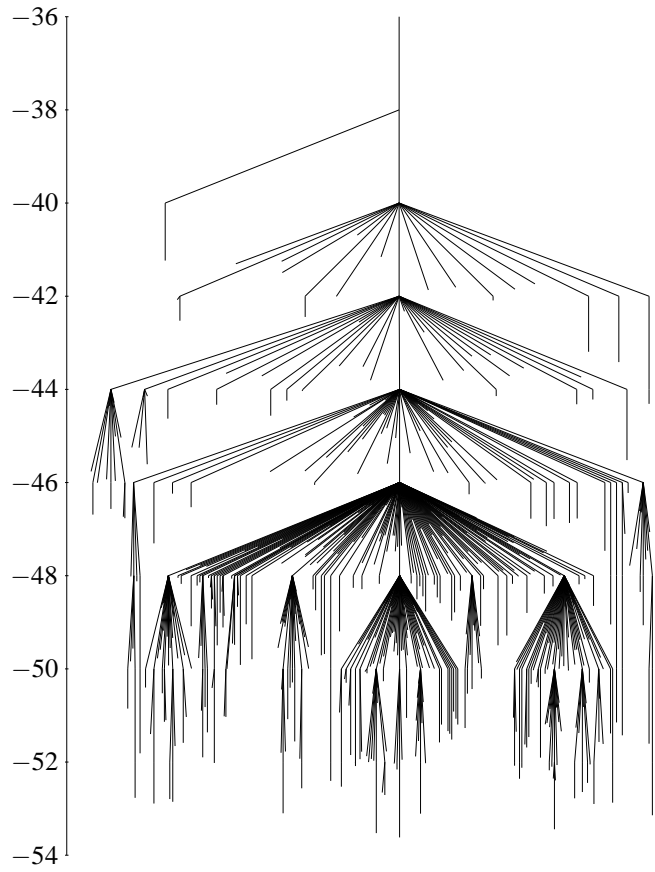}}
\caption{
Disconnectivity graph for the BLN model, based on a sample of 500 minima
and 636 transition states. The energy is in units of the parameter $\epsilon$.
}
\label{fig:treep46}
\end{figure}

The disconnectivity graph for the low-energy regions of the BLN model
landscape is shown in \fig{treep46}, using the sample of 500 minima and 636
transition states obtained in \sect{explore}.
It is immediately apparent that the PES is not a single funnel.
In fact, it is a good example of a rough energy landscape, with repeated
splitting at successive nodes and long descending branches. A number of low-energy
structures exist which are separated by high barriers. Even if the barriers
were not so high, there would be little thermodynamic driving force towards the
global minimum. The fact that the attractive forces are of relatively long range
and non-specific character means that it is possible to construct many significantly
different structures from common motifs such as the four strands in the global minimum.
For example, some of the low-energy minima differ only by the relative positions of the
two purely hydrophobic strands. These can register with each other in a number of
positions, related visually by a parallel slide. However, such a slide would be
an unlikely mechanism because all the non-bonded interactions would be disrupted
at once. Instead, the shortest path between such structures typically proceeds
through over ten separate rearrangements.
\par
Other ways in which low-energy structures are related involve a reorientation
of the hydrophobic strands, so that the beads which are outermost and those
that come into contact in the core in \fig{p46} are interchanged. Again, such
a process involves many steps and a high barrier. The neutral turn regions
can also adopt a number of configurations. The barriers between structures
related in this way tend to be somewhat smaller, since the torsion potential
is weaker in these regions.
\par
The same structural database that is used to construct the disconnectivity graph
can also be analysed in terms of `monotonic sequences' of connected
minima in which the potential energy decreases with every step.\cite{Berry95a,Kunz95a}
The collection of sequences leading to a
particular minimum define what we will call a monotonic sequence basin (MSB).
Whilst the super-basin of the disconnectivity graph is defined at a specified energy,
a monotonic sequence basin is a fixed feature of the landscape.
\par
Berry et al.\cite{Berry97a} have characterized some monotonic sequences leading
to the global minimum of the BLN model. The sequences are connected by barriers
that are relatively low compared with the energy gradient along the sequence,
leading these authors to place the BLN model into the category of `structure seekers'.
We note, however, that only 67 of our sample of 500 minima lie on monotonic
sequences to the global minimum, so that such sequences are not representative
of paths to the global minimum. Furthermore, other low-energy minima also lie
at the bottom of separate monotonic sequences of comparable or even larger sets
of minima. Hence, this system `seeks' only a general $\beta$-barrel structure;
consideration of the arrangement of the monotonic sequences shows that significantly
different low-energy minima will be reached from different starting configurations,
and interconversion of these minima will be relatively slow with little preference
for any given one.

\subsection{The Effects of Frustration}

The folding characteristics of the BLN model have recently been questioned in other
studies. Guo and Brooks\cite{Guo97a} used MD simulations and a histogram
method to study the thermodynamics of folding. They identified a collapse transition
to compact states with a peak in the specific heat, and a folding transition in
terms of a similarity parameter with the global minimum. The free energy surface as
a function of this parameter and the compactness showed that collapse occurs before
any appreciable native structure is attained, rather than the cooperative collapse
and structuring expected for a good folder. Nymeyer et al.\cite{Nymeyer98a} inferred
the roughness of the energy landscape from the model's thermodynamic and dynamic
behaviour.\cite{potentialnote}
To demonstrate the effects of frustration, they compared their simulations of the BLN
model with a modified version in which the frustration is largely eliminated. We now
characterize the energy landscape of this modified model.
\par
To remove the effects of frustration in the BLN model, all attractive interactions
between pairs of monomers that are not in contact in the native state (global minimum)
are removed. This transformation is equivalent to setting $D_{ij}=0$ in
\eq{barrelpot} for non-bonded pairs of hydrophobic monomers which are separated by
more than $1.167\,\sigma$ in the global minimum. This change increases the heterogeneity of
the interactions, since it makes the attractive forces more specific. The modified
potential was termed `\Go-like', following \Go and collaborators, who constructed
model lattice proteins by defining attractive interactions between neighbouring
non-bonded monomers in an assumed ground state structure.\cite{Ueda78a}
\par
\begin{figure}[ht]
\psfrag{-07}[r][r]{\fsz $-7$}
\psfrag{-09}[r][r]{\fsz $-9$}
\psfrag{-11}[r][r]{\fsz $-11$}
\psfrag{-13}[r][r]{\fsz $-13$}
\psfrag{-15}[r][r]{\fsz $-15$}
\psfrag{-17}[r][r]{\fsz $-17$}
\psfrag{-19}[r][r]{\fsz $-19$}
\psfrag{-21}[r][r]{\fsz $-21$}
\psfrag{-23}[r][r]{\fsz $-23$}
\psfrag{-25}[r][r]{\fsz $-25$}
\psfrag{-27}[r][r]{\fsz $-27$}
\psfrag{-29}[r][r]{\fsz $-29$}
\psfrag{-31}[r][r]{\fsz $-31$}
\centerline{\includegraphics[width=85mm]{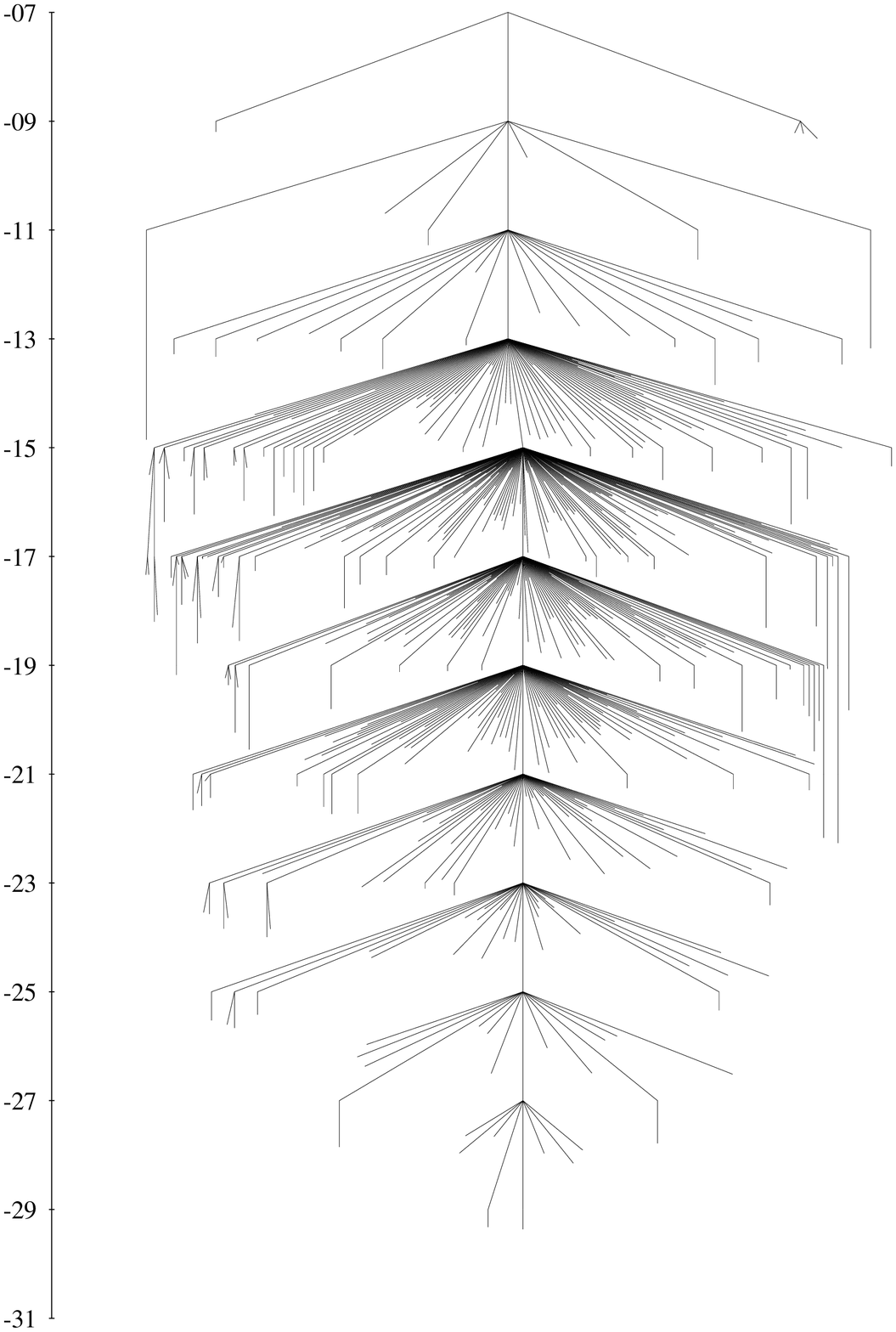}}
\caption{
Disconnectivity graph for the \Go-like model, based on a sample of 500 minima
and 805 transition states. The energy is in units of the parameter $\epsilon$.
}
\label{fig:treeGo}
\end{figure}

Performing a survey of the energy landscape of the \Go-like model as for the BLN
model above produced 805 transition states linking the 500 low-lying minima. The
disconnectivity graph is shown in \fig{treeGo}. The appearance is much more
funnel-like, with no low-energy minima separated from the global minimum by
large barriers. Relaxing the BLN global minimum with the \Go-like potential
actually produces the second-lowest
energy structure; a similar structure differing in the
orientation of one of the turns lies slightly lower. The energy range of the
disconnectivity graph is a much larger proportion of the global minimum well depth
than in the analogous graph for the BLN model (\fig{treep46}). This range reflects the
lower density of minima per unit energy in the \Go-like system that results from
the specificity of the attractive forces. The highest-energy minima in the BLN
sample were still relatively compact, whereas those for the \Go-like model showed
considerable unfolding of the $\beta$-barrel.
\par
\begin{figure}[ht]
\psfrag{(a)}[][]{}
\psfrag{(b)}[][]{}
\psfrag{energy / e}[t][]{\fsz $\text{energy}/\epsilon$}
\psfrag{Sgmin / s}[][]{\fsz $S_{\rm gmin}/\sigma$}
\psfrag{0}{\ssz 0}
\psfrag{100}{\ssz 100}
\psfrag{200}{\ssz 200}
\psfrag{300}{\ssz 300}
\psfrag{400}{\ssz 400}
\psfrag{-10}{\ssz $-10$}
\psfrag{-20}{\ssz $-20$}
\psfrag{-30}{\ssz $-30$}
\psfrag{-40}{\ssz $-40$}
\psfrag{-45}{\ssz $-45$}
\psfrag{-50}{\ssz $-50$}
\psfrag{-55}{\ssz $-55$}
\centerline{\includegraphics[width=85mm]{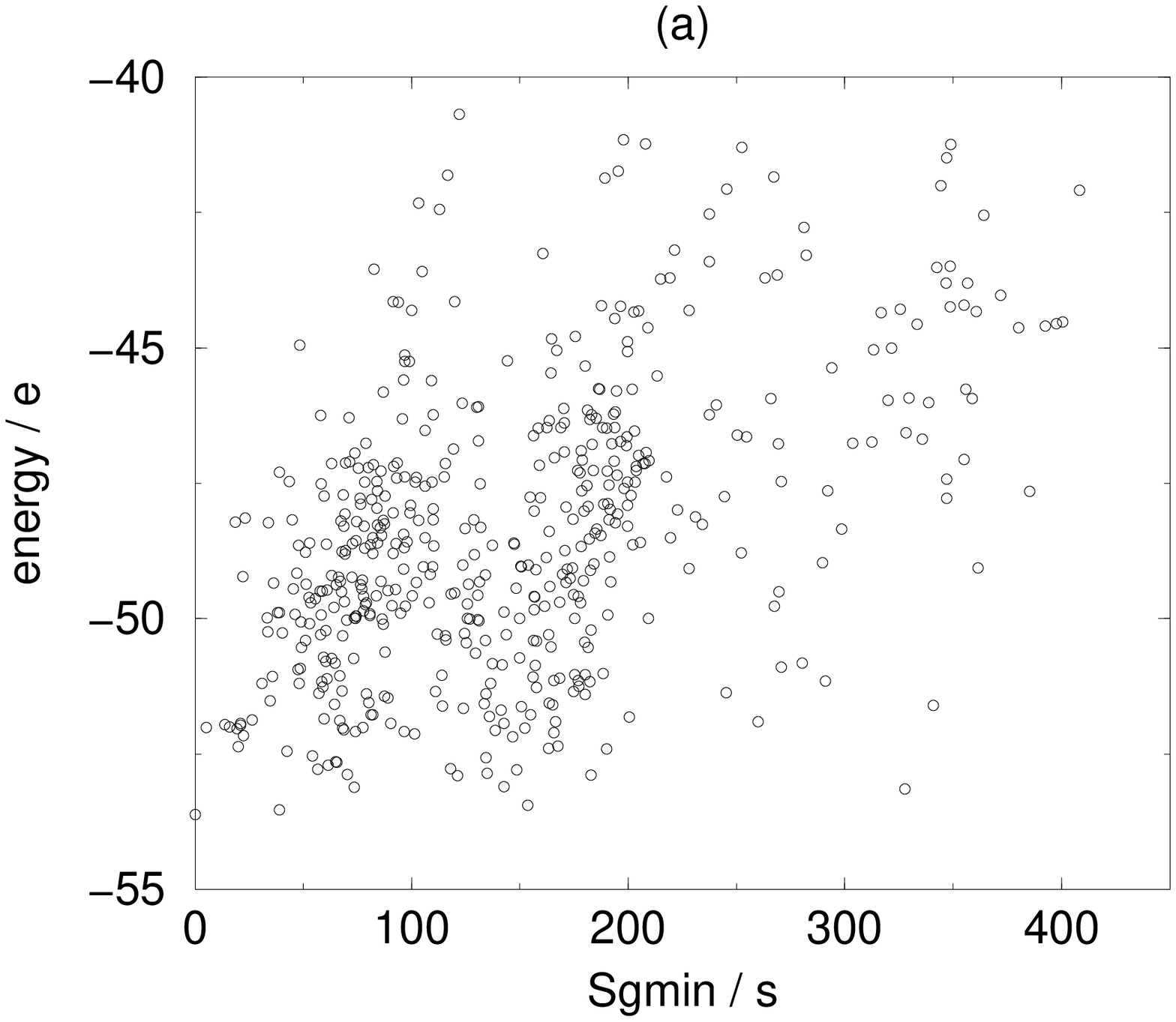}}
\centerline{\includegraphics[width=85mm]{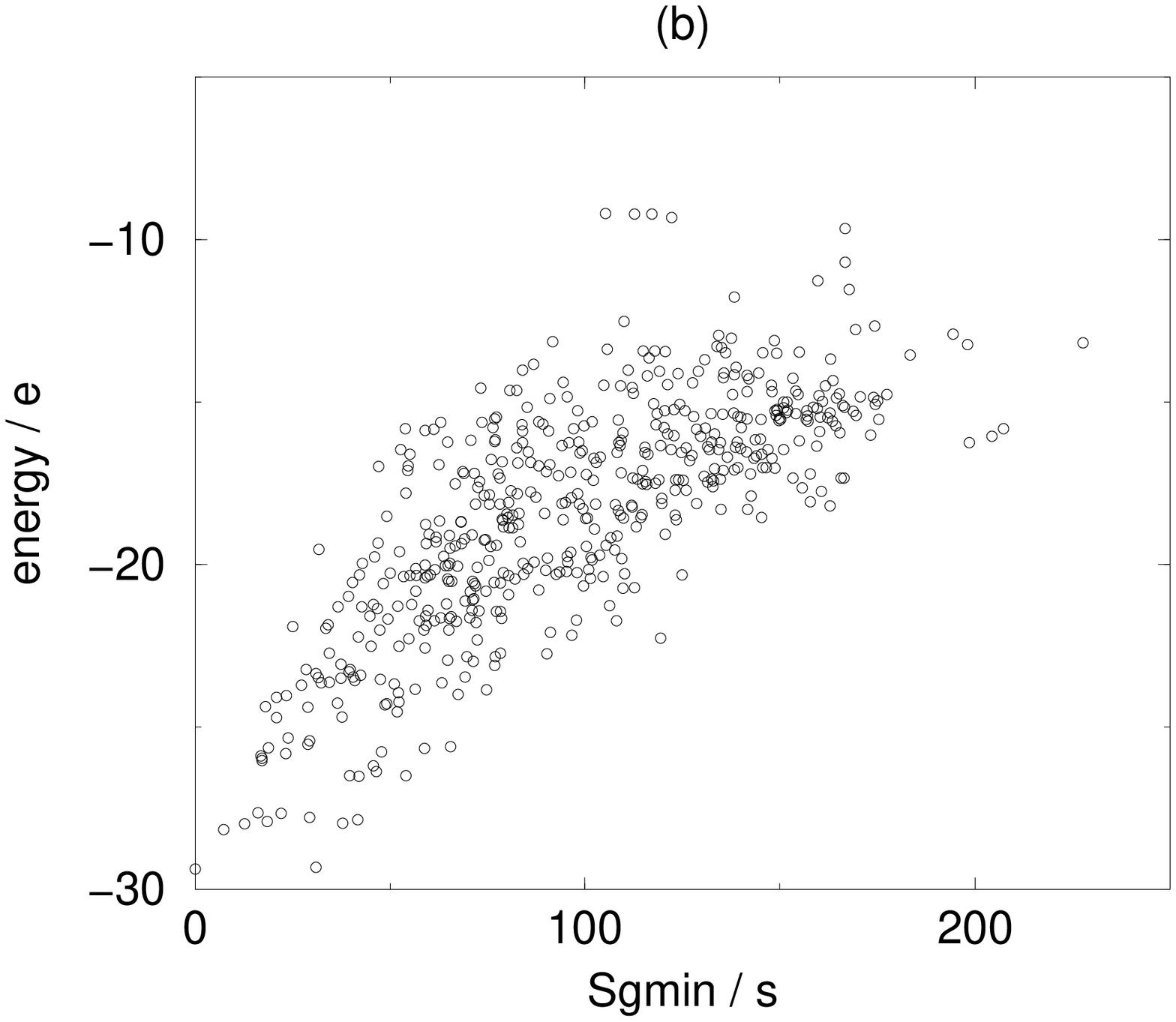}}
\caption{
Energy of minima as a function of the integrated path length along the shortest
path to the global minimum. Upper panel: the BLN model; lower panel: the \Go-like model.
}
\label{fig:sgminp46}
\end{figure}
\begin{figure}[ht]
\psfrag{number of minima}[t][]{\fsz number of minima}
\psfrag{ngm}[t][]{\fsz number of steps}
\psfrag{0}{\ssz 0}
\psfrag{20}{\ssz 20}
\psfrag{40}{\ssz 40}
\psfrag{60}{\ssz 60}
\psfrag{80}{\ssz 80}
\psfrag{100}{\ssz 100}
\psfrag{5}{\ssz 5}
\psfrag{10}{\ssz 10}
\psfrag{15}{\ssz 15}
\psfrag{25}{\ssz 25}
\centerline{\includegraphics[width=85mm]{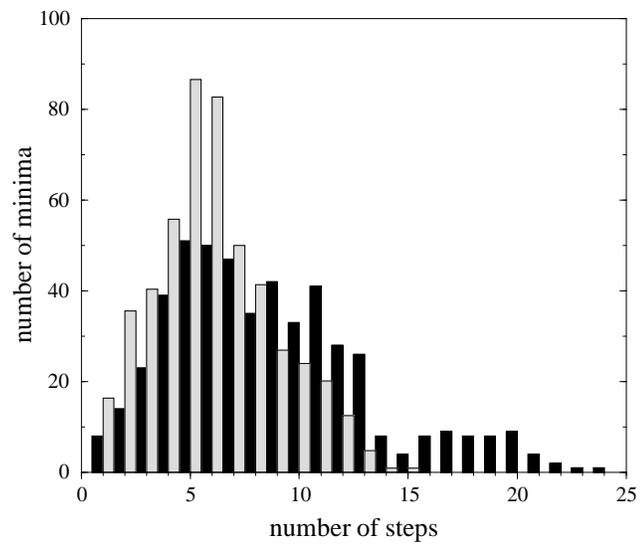}}
\caption{
Distribution of the number of rearrangements along the shortest path from a given
minimum to the global minimum for the BLN model (black) and the \Go-like model
(grey).
}
\label{fig:ngmbarrel}
\end{figure}

The plots of energy versus shortest integrated path length to the global minimum
in \fig{sgminp46} display the difference between the BLN and \Go-like energy
landscapes clearly. For the BLN model there is little correlation between distance and
energy, whereas for the \Go-like model the energy rises with distance, as one would
expect in a funnel-like landscape.\cite{LJpaper} The number of
individual rearrangements along the shortest paths to the global minimum is shown
for both models in \fig{ngmbarrel}. The distribution for the BLN model is broader,
with some minima lying as far as 24 steps from the global minimum, in contrast with
a maximum of 15 for the \Go-like model. This reveals the greater organization of
the \Go-like energy landscape into pathways converging at the global minimum.
\par
A funnel-like interpretation for the \Go-like model is also encouraged by the changes
in the average properties of the individual paths between minima, as demonstrated
in \tab{p46paths}. Uphill barriers are, on average, higher and downhill barriers
lower for the \Go-like model, producing a steeper downhill gradient between minima.
However, the funnel of the \Go-like model is far from ideal. A monotonic sequence
analysis shows that only 124 of the 500 minima lie in the primary MSB, so that the
relaxation from an arbitrary structure to the global minimum is likely to involve
a number of uphill steps.
\par
In simulations, Nymeyer et al.\cite{Nymeyer98a} found that the collapse from
unfolded states and the formation of native structure occurred cooperatively for
the \Go-like model, producing a single narrow peak in the heat capacity. They
also showed that glassy dynamics, as measured by non-exponential relaxation
from unfolded states, starts at temperatures just below the collapse for the
BLN model, hindering the search for the native structure. In the \Go-like
model, in contrast, glassy dynamics only set in below the folding temperature,
where the global minimum still has a large equilibrium probability. These
results are entirely in accord with those expected from the direct characterization
of the energy landscape presented here.

\section{Conclusions}

The disconnectivity graph analysis of the 3-colour, 46-bead model polypeptide
reveals a frustrated energy landscape with a number of low-lying $\beta$-barrel 
structures in competition with the global potential energy minimum. Although
relaxation to one of these $\beta$-barrel minima may be quite efficient, much longer
time scales are needed for the system to reliably locate the global minimum,
in agreement with previous simulations.
\par
In contrast, when the frustration is removed by changing the potential to
a \Go-type model, the landscape is transformed to one where the global minimum
should be located easily. The competitive low-lying minima disappear
following the transformation, and the metastable minima are organised with an
energy gradient towards the global minimum. Our results illustrate the utility of the
disconnectivity graph approach as a tool to rationalize and predict structural,
dynamic and thermodynamic behaviour from the potential energy surface.

\section*{Acknowledgements}
We are grateful to the Royal Society and the EPSRC for supporting this research,
and to Dr John Rose for providing his Fortran routines for the model potential.

\newpage
\bibliographystyle{thesis}

\begin{thebibliography}{10}

\bibitem{Toulouse77a}
G.~Toulouse, \cop~\textbf{2}, 115 (1977).

\bibitem{Bryngelson95a}
J.~D. Bryngelson, J.~N. Onuchic, N.~D. Socci  and P.~G. Wolynes,
  \psfg~\textbf{21}, 167 (1995).

\bibitem{Stillinger99a}
F.~H. Stillinger, \pre~\textbf{59}, 48 (1999).

\bibitem{Levinthal69a}
C.~Levinthal, in \emph{M\"ossbauer spectroscopy in biological systems,
  proceedings of a meeting held at Allerton House, Monticello, Illinois},
  edited by P.~DeBrunner, J.~Tsibris  and E.~Munck, p.~22, Urbana (1969),
  University of Illinois Press.

\bibitem{Levinthal68a}
C.~Levinthal, \jchp~\textbf{65}, 44 (1968).

\bibitem{Camacho93a}
C.~J. Camacho and D.~Thirumalai, \prl~\textbf{71}, 2505 (1993).

\bibitem{Sali94a}
A.~\u{S}ali, E.~Shakhnovich  and M.~Karplus, \nat~\textbf{369}, 248 (1994).

\bibitem{Socci94a}
N.~D. Socci and J.~N. Onuchic, \jcp~\textbf{101}, 1519 (1994).

\bibitem{Zwanzig92a}
R.~Zwanzig, A.~Szabo  and B.~Bagchi, \pnasu~\textbf{89}, 20 (1992).

\bibitem{Zwanzig95a}
R.~Zwanzig, \pnasu~\textbf{92}, 9801 (1995).

\bibitem{Leopold92a}
P.~E. Leopold, M.~Montal  and J.~N. Onuchic, \pnasu~\textbf{89}, 8721 (1992).

\bibitem{Wolynes95a}
P.~G. Wolynes, J.~N. Onuchic  and D.~Thirumalai, \sci~\textbf{267}, 1619
  (1995).

\bibitem{Socci96a}
N.~D. Socci, J.~N. Onuchic  and P.~G. Wolynes, \jcp~\textbf{104}, 5860 (1996).

\bibitem{Wales98c}
D.~J. Wales, M.~A. Miller  and T.~R. Walsh, \nat~\textbf{394}, 758 (1998).

\bibitem{Doye99b}
J.~P.~K. Doye, M.~A. Miller  and D.~J. Wales, \jcp~\textbf{110}, 6896 (1999).

\bibitem{Becker97a}
O.~M. Becker and M.~Karplus, \jcp~\textbf{106}, 1495 (1997).

\bibitem{Czerminski90a}
R.~Czerminski and R.~Elber, \jcp~\textbf{92}, 5580 (1990).

\bibitem{Levy98a}
Y.~Levy and O.~M. Becker, \prl~\textbf{81}, 1126 (1998).

\bibitem{Berry97a}
R.~S. Berry, N.~Elmaci, J.~P. Rose  and B.~Vekhter, \pnasu~\textbf{94}, 9520
  (1997).

\bibitem{Honeycutt90a}
J.~D. Honeycutt and D.~Thirumalai, \pnasu~\textbf{87}, 3526 (1990).

\bibitem{Honeycutt92a}
J.~D. Honeycutt and D.~Thirumalai, \bp~\textbf{32}, 695 (1992).

\bibitem{Vekhter99a}
B.~Vekhter and R.~S. Berry, \jcp~\textbf{110}, 2195 (1999).

\bibitem{Murrell68a}
J.~N. Murrell and K.~J. Laidler, \jcsft~\textbf{64}, 371 (1968).

\bibitem{Pancir74a}
J.~Panc{\'\i}\v{r}, \cccc~\textbf{40}, 1112 (1974).

\bibitem{Cerjan81a}
C.~J. Cerjan and W.~H. Miller, \jcp~\textbf{75}, 2800 (1981).

\bibitem{Simons83a}
J.~Simons, P.~J{\o}rgenson, H.~Taylor  and J.~Ozment, \jpc~\textbf{87}, 2745
  (1983).

\bibitem{Banerjee85a}
A.~Banerjee, N.~Adams, J.~Simons  and R.~Shepard, \jpc~\textbf{89}, 52 (1985).

\bibitem{Baker86a}
J.~Baker, \jcc~\textbf{7}, 385 (1986).

\bibitem{Wales94a}
D.~J. Wales, \jcp~\textbf{101}, 3750 (1994).

\bibitem{Wales96c}
D.~J. Wales and T.~R. Walsh, \jcp~\textbf{105}, 6957 (1996).

\bibitem{Tsai93b}
C.~J. Tsai and K.~D. Jordan, \jpc~\textbf{97}, 11227 (1993).

\bibitem{Doye97a}
J.~P.~K. Doye and D.~J. Wales, \zpd~\textbf{40}, 194 (1997).

\bibitem{Barkema96a}
G.~T. Barkema and N.~Mousseau, \prl~\textbf{77}, 4358 (1996).

\bibitem{Mousseau98a}
N.~Mousseau and G.~T. Barkema, \pre~\textbf{57}, 2419 (1998).

\bibitem{Stillinger82a}
F.~H. Stillinger and T.~A. Weber, \pra~\textbf{25}, 978 (1982).

\bibitem{Berry95a}
R.~S. Berry and R.~Breitengraser-Kunz, \prl~\textbf{74}, 3951 (1995).

\bibitem{Kunz95a}
R.~E. Kunz and R.~S. Berry, \jcp~\textbf{103}, 1904 (1995).

\bibitem{Guo97a}
Z.~Guo and C.~L. Brooks~III, \bp~\textbf{42}, 745 (1997).

\bibitem{Nymeyer98a}
H.~Nymeyer, A.~E. Garc\'{\i}a  and J.~N. Onuchic, \pnasu~\textbf{95}, 5921
  (1998).

\bibitem{potentialnote}
There is some discrepancy in the details of the potential. In
  \paper{Nymeyer98a} the bond angle term of \eq{barrelpot} is defined without
  the factor of $1/2$, but the value of $K_\theta$ specified is {\em twice}
  that in \paper{Honeycutt92a}. Furthermore, \paper{Nymeyer98a} specifies
  $A_i=1.2$ and $B_i=0.2$ for the dihedral term in \eq{barrelpot} when no more
  than one N monomer is involved, rather than $A_i=B_i=1.2$. This second
  discrepancy is almost certainly a typographical error, since only the trans
  conformation would then be locally stable. In any case, the thermodynamic
  results are in fairly good agreement with those of \paper{Guo97a}.

\bibitem{Ueda78a}
Y.~Ueda, H.~Taketomi  and N.~G\=o, \bp~\textbf{17}, 1531 (1978).

\bibitem{LJpaper}
J.~P.~K. Doye, M.~A. Miller  and D.~J. Wales, \jcp~\textbf{000}, submitted
  (1999).

\end{thebibliography}

\newcommand\aciee{Angew. Chem. Int. Ed. Engl.\xspace}
\newcommand\ac{Acta. Crystallogr.\xspace}
\newcommand\acp{Adv. Chem. Phys.\xspace}
\newcommand\acr{Acc. Chem. Res.\xspace}
\newcommand\ajp{Am. J. Phys.\xspace}
\newcommand\ap{Ann. Physik\xspace}
\newcommand\arpc{Ann. Rev. Phys. Chem.\xspace}
\newcommand\bbpc{Ber. Bunsenges. Phys. Chem.\xspace}
\newcommand\bc{Biochemistry\xspace}
\newcommand\bmk{Biometrika\xspace}
\newcommand\bp{Biopolymers\xspace}
\newcommand\cccc{Coll. Czech. Chem. Comm.\xspace}
\newcommand\cop{Comm. Phys.\xspace}
\newcommand\cp{Chem. Phys.\xspace}
\newcommand\cpc{Comp. Phys. Comm.\xspace}
\newcommand\cpl{Chem. Phys. Lett.\xspace}
\newcommand\crev{Chem. Rev.\xspace}
\newcommand\ea{Electrochim. Acta\xspace}
\newcommand\el{Europhys. Lett.\xspace}
\newcommand\epjd{Eur. Phys. J. D\xspace}
\newcommand\fd{Faraday Discuss.\xspace}
\newcommand\ic{Inorg. Chem.\xspace}
\newcommand\ijmpc{Int. J. Mod. Phys. C\xspace}
\newcommand\ijqc{Int. J. Quant. Chem.\xspace}
\newcommand\jcis{J. Colloid Interface Sci.\xspace}
\newcommand\jcsft{J. Chem. Soc., Faraday Trans.\xspace}
\newcommand\jacers{J. Am. Ceram. Soc.\xspace}
\newcommand\jacs{J. Am. Chem. Soc.\xspace}
\newcommand\jas{J. Atmos. Sci.\xspace}
\newcommand\jcc{J. Comp. Chem.\xspace}
\newcommand\jchp{J. Chim. Phys.\xspace}
\newcommand\jcp{J. Chem. Phys.\xspace}
\newcommand\jce{J. Chem. Ed.\xspace}
\newcommand\jcscc{J. Chem. Soc., Chem. Commun.\xspace}
\newcommand\jetp{J. Exp. Theor. Phys. (Russia)\xspace}
\newcommand\jmb{J. Mol. Biol.\xspace}
\newcommand\jmsp{J. Mol. Spec.\xspace}
\newcommand\jmst{J. Mol. Struct.\xspace}
\newcommand\jncs{J. Non-Cryst. Solids\xspace}
\newcommand\jpa{J. Phys. A\xspace}
\newcommand\jpb{J. Phys. B\xspace}
\newcommand\jpc{J. Phys. Chem.\xspace}
\newcommand\jpca{J. Phys. Chem. A\xspace}
\newcommand\jpcb{J. Phys. Chem. B\xspace}
\newcommand\jpcm{J. Phys. Condensed Matter.\xspace}
\newcommand\jpcs{J. Phys. Chem. Solids.\xspace}
\newcommand\jpsj{J. Phys. Soc. Jpn.\xspace}
\newcommand\jsp{J. Stat. Phys.\xspace}
\newcommand\jvsta{J. Vac. Sci. Technol. A\xspace}
\newcommand\mg{Math. Gazette\xspace}
\newcommand\molp{Mol. Phys.\xspace}
\newcommand\nat{Nature\xspace}
\newcommand\nsb{Nature Struct. Biol.\xspace}
\newcommand\pac{Pure. Appl. Chem.\xspace}
\newcommand\pd{Physica D\xspace}
\newcommand\phys{Physics\xspace}
\newcommand\pma{Philos. Mag. A\xspace}
\newcommand\pmagb{Philos. Mag. B\xspace}
\newcommand\pnasu{Proc. Natl. Acad. Sci. USA\xspace}
\newcommand\ppmsj{Proc. Phys.-Math. Soc. Japan\xspace}
\newcommand\pr{Phys. Rev.\xspace}
\newcommand\prep{Phys. Reports\xspace}
\newcommand\pra{Phys. Rev. A\xspace}
\newcommand\prb{Phys. Rev. B\xspace}
\newcommand\prc{Phys. Rev. C\xspace}
\newcommand\prd{Phys. Rev. D\xspace}
\newcommand\pre{Phys. Rev. E\xspace}
\newcommand\prl{Phys. Rev. Lett.\xspace}
\newcommand\prsa{Proc. R. Soc. A\xspace}
\newcommand\psfg{Proteins: Struct., Func. and Gen.\xspace}
\newcommand\ptps{Prog. Theor. Phys. Supp.\xspace}
\newcommand\rmp{Rev. Mod. Phys.\xspace}
\newcommand\sci{Science\xspace}
\newcommand\sa{Sci. Amer.\xspace}
\newcommand\ssci{Surf. Sci.\xspace}
\newcommand\tca{Theor. Chim. Acta\xspace}
\newcommand\zpb{Z. Phys. B.\xspace}
\newcommand\zpc{Z. Phys. Chem.\xspace}
\newcommand\zpd{Z. Phys. D\xspace}

\clearpage
\newpage
\section*{Tables}

\begin{table}[ht]
\caption{
Properties of individual pathways for the BLN and \Go-like models.
$b^{\rm up}_i$ is the larger (uphill) barrier height
between the two minima connected by transition state $i$, and
$b^{\rm down}_i$ is the smaller (downhill) barrier.
$\Delta E^{\rm con}_i=b^{\rm up}_i-b^{\rm down}_i$ is the energy difference
between the two minima. The angle brackets indicate averaging over the sample
of pathways. The units of energy are $\epsilon$.
}
\begin{center}
\begin{tabular}{lll}
\toprule
Model & BLN & \Go-like \\
\midrule
$\langle b^{\rm up}\rangle_{\rm p}$        & 2.59  & 3.07 \\
$\langle b^{\rm down}\rangle_{\rm p}$      & 0.862 & 0.635 \\
$\langle\Delta E^{\rm con}\rangle_{\rm p}$ & 1.73  & 2.43 \\
\bottomrule
\end{tabular}
\end{center}
\label{tab:p46paths}
\end{table}

\clearpage
\newpage
\section*{Figures}

\end{document}